\documentclass[twocolumn,aps,prl,amsmath,showpacs,amssymb,floatfix]{revtex4}
\usepackage{tabularx}
\usepackage{bm}
\usepackage{subfigure}
\usepackage{euscript}
\usepackage{graphicx}
\usepackage{color}
\usepackage[colorlinks=true,linkcolor=blue]{hyperref}%
\usepackage{amsfonts}
\usepackage{exscale}
\usepackage{amsbsy}

\begin{document}

\title{Topological phases, topological flat bands, and topological excitations \\ in a one-dimensional dimerized lattice with spin-orbit coupling}

\author{Zhongbo Yan}
\author{Shaolong Wan}
\email[]{slwan@ustc.edu.cn}
\affiliation{Institute for Theoretical
Physics and Department of Modern Physics University of Science and
Technology of China, Hefei, 230026, P. R. China}
\date{\today}

\begin{abstract}
The Su-Schrieffer-Heeger (SSH) model describes a one-dimensional
$Z_{2}$ topological insulator, which has two topological distinct
phases corresponding to two different dimerizations. When
spin-orbit coupling is introduced into the SSH model, we find the
structure of the Bloch bands can be greatly changed, and most
interestingly, a new topological phase with single zero-energy
bound state which exhibits non-Abelian statistics at each end
emerges, which suggests that a new topological invariant is needed
to fully classify all phases. In a comparatively large range of
parameters, we find that spin-orbit coupling induces completely
flat band with nontrivial topology. For the case with non-uniform
dimerizaton, we find that spin-orbit coupling changes the
symmetrical structure of topological excitations known as solitons
and antisolitons and when spin-orbit coupling is strong enough to
induce a topological phase transition, the whole system is
topologically nontrivial but with the disappearance of solitons
and antisolitons, consequently, the system is a real topological
insulator with well-protected end states.
\end{abstract}


\maketitle

{\it Introduction.---}Since the discovery of integer quantum Hall
effect \cite{K. Klitzing}, the concept of topology has become
increasingly popular and important in condensed matter physics
\cite{D. Thouless}. Topological phases, usually characterized by
topological invariants which are connected to the energy spectrum
and the nature of the wave function \cite{Xiao-Gang Wen, X.-G.
Wen}, can appear in different dimensional systems with higher or
lower symmetries \cite{A.P. Schnyder, A. Y. Kitaev}, and due to
the robustness against disorder, they have the appealing potential
application in topological quantum computation \cite{A. Kitaev}.

In the past few years, the theoretical predictions \cite{C. L.
Kane, B. A. Bernevig} and the experimental observations \cite{M.
Konig, D. Hsieh} of topological insulator have stimulated strong
and continuous interest in predicting new classes of materials
with nontrivial topological properties. From the lessons of
topological insulators, we have learned that spin-orbit coupling
is usually a natural ingredient to generate topological phases,
$e.g.$ topological superconductors which host desirable Majorana
fermions \cite{Roman M. Lutchyn, J. Alicea}. Inspired by the
recent experimental realization of direct measurement of Zak phase
in one-dimensional topological Bloch bands \cite{M. Atala}, and
the realization of spin-orbit coupling in a one-dimensional cold
atomic system \cite{Y.-J. Lin, P. J. Wang}, in this work we
investigate the influence of spin-orbit coupling on the topology
of Bloch bands in the one-dimensional double-well optical lattice
used in the experiment \cite{M. Atala}.

The one-dimensional double-well optical lattice is a cold atomic
realization of the well-known Su-Schrieffer-Heeger (SSH) model
\cite{W. P. Su}. The SSH model describes a  $Z_{2}$ topological
insulator, which has two topological distinct phases corresponding
to two different dimerizations. In one dimension, Zak phase
determines the topology of Bloch bands \cite{J. Zak}, and it takes
value either $-\pi/2$ or $\pi/2$ (here we follow the choice of
unit cell in Ref.\cite{M. Atala}, a different choice of unit cell
gives the usual value $0$ or $\pi$), and therefore, it is usually
taken as the $Z_{2}$ invariant to characterize the two topological
distinct phases, and a change of Zak phase indicates a topological
phase transition. After introducing spin-orbit coupling, the
extended SSH model still be a very simple model. However,
spin-orbit coupling will lift the original degeneracy of spin
degree and induce strong influence on the structure of Bloch
bands, and these consequently greatly rich the physics of the
system. The main results induced by spin-orbit coupling include:
(i) a series of topological phase transitions; (ii) the emergence
of a new topological phase with single zero-energy bound state
which exhibits non-Abelian statistics at each end; (iii) the
formation of completely flat band with nontrivial topology; (iv)
the change of the symmetric form of solitons and antisolitons; (v)
the disappearance and the reappearance of solitons and
antisolitons with the increase of the strength and the asymmetry
of spin-orbit coupling.

{\it Theoretical model.---}We consider the one-dimensional double-well lattice
realized in the experiment \cite{M. Atala}, but with the introduction of spin-orbit
coupling. The Hamiltonian describing the system is given by
\begin{eqnarray}
\hat{H} &=& -\sum_{i,\sigma} \left(J \hat{a}_{i,\sigma}^{\dag}
\hat{b}_{i,\sigma} + J^{'} \hat{a}_{i,\sigma}^{\dag} \hat{b}_{i-1,\sigma} + h.c.\right) \nonumber\\
&&+ \sum_{i,\sigma} \left(\lambda \hat{a}_{i,\sigma}^{\dag}
\hat{b}_{i,-\sigma} - \lambda^{'} \hat{a}_{i,\sigma}^{\dag}
\hat{b}_{i-1,-\sigma} +h.c. \right), \label{1}
\end{eqnarray}
where $J,~J^{'}$ denote modulated tunneling amplitudes within the
unit cell, $\hat{a}_{i,\sigma}^{\dag}(\hat{b}_{i,\sigma}^{\dag})$
are the particle creation operators for an atom with spin $\sigma$
($\uparrow$ or $\downarrow$) in the $i$th lattice cell.
$\lambda,~\lambda^{'}$ denote modulated spin-orbit coupling. For
the special case, $\lambda = \lambda^{'}$, the spin-orbit coupling
is just the form that can be realized in current cold-atomic
experiments \cite{V. Galitski}.

In the absence of spin-orbit coupling, $i.e.$ $\lambda =
\lambda^{'} =0 $, the above Hamiltonian corresponds to the
well-known Su-Schrieffer-Heeger model (SSH) of polyacetylene
\cite{W. P. Su}. In the SSH model, spin degrees are decoupled and
the Bloch bands for spin-up and spin-down are degenerate. For each
spin degree, the topology of the Bloch bands is classified by the
$Z_{2}$ invariant known as Zak phase. As the bands are degenerate,
then if the lower band (the occupied band) is of nontrivial
topology for one spin degree, the lower band for the other spin
degree is also topologically nontrivial, this implies that the
number of bound states for each end is even. Due to the
degeneracy, the upper Bloch band and the lower Bloch band for
spin-up and spin-down are simultaneously touched at the
topological phase transition point $J = J^{'}$, as a result, the
change of the Zak phase accompanying with the topological phase
transition is $2 \pi$, instead of $\pi$.

After the appearance of spin-orbit coupling, the spin is no longer
a good quantum number, and the degeneracy of the Bloch bands will
be lifted, then by tuning the strength and asymmetry of the
modulated spin-orbit coupling, only one of the lower band
corresponding to one of the two helicities will undergo the
topologically trivial to non-trivial or non-trivial to trivial
transition, and the other lower band will keep its topology. This
indicates that if the system is of nontrivial topology in the
absence of spin-orbit coupling, then even the spin-orbit coupling
drives a topological phase transition, the system is still
topologically nontrivial, but with a change of the number of end
bound states. This suggests that using a $Z_{2}$ invariant can no
longer fully characterize all phases, and we need to introduce a
new topological invariant. In the following, we will give a detail
investigation of the influence of spin-orbit coupling.

The Hamiltonian (\ref{1}) in momentum space takes the form,
\begin{eqnarray}
\hat{H} = \sum_{k} \hat{h}(k)
&=& \sum_{k} \left\{-\left[(\text{Re} \rho_{k}) \sigma_{x} - (\text{Im} \rho_{k}) \sigma_{y} \right] \right. \nonumber\\
&&+ \left. \left[(\text{Re} \delta_{k}) \sigma_{x} - (\text{Im}
\delta_{k}) \sigma_{y} \right] \tau_{x} \right\}, \label{2}
\end{eqnarray}
where $\sigma_{x},\sigma_{y}$ are the Pauli matrices for
sublattice and $\tau_{x}$ is the Pauli matrix for spin, and
\begin{eqnarray}
\rho_{k} &=& J e^{ikd/2} + J^{'} e^{-ikd/2} = |\epsilon_{1}(k)| e^{i\theta_{1}(k)}, \nonumber\\
\delta_{k} &=& \lambda e^{ikd/2} - \lambda^{'} e^{-ikd/2} =
|\epsilon_{2}(k)| e^{i\theta_{2}(k)}. \label{3}
\end{eqnarray}
The Hamiltonian obviously has particle-hole symmetry and chiral symmetry,
\begin{eqnarray}
-\hat{h}(k) &=& \sigma_{z} \hat{h}^{T}(-k) \sigma_{z}, \nonumber\\
-\hat{h}(k) &=& \sigma_{z} \hat{h}(k) \sigma_{z}. \label{4}
\end{eqnarray}
As $\sigma_{z}\sigma_{z}^{\dag}=1$ and
$\sigma_{z}^{T}=\sigma_{z}$, then according to Ref.\cite{A.P.
Schnyder}, we know the Hamiltonian belongs to the symmetry class
$BDI$ (chiral orthogonal), and the phases are classified by $Z$,
an integer. In fact, the SSH model also belongs to this symmetry
class, however, due to the degeneracy of the Bloch bands, a
$Z_{2}$ invariant can fully classify the phases.

To determine the topology of the system, we need to determine the
topology of the occupied Bloch bands, which can be realized by
calculating the Zak phase of the corresponding band,
\begin{eqnarray}
\varphi_{Zak} = i \sum_{\sigma} \int_{-G/2}^{G/2}
(\alpha_{k,\sigma}^{*} \partial_{k} \alpha_{k,\sigma} +
\beta_{k,\sigma}^{*} \partial_{k} \beta_{k,\sigma}) d k, \label{5}
\end{eqnarray}
where $\alpha_{k,\sigma}$ and $\beta_{k,\sigma}$ are the four
components of a spinor $\mathbf{u}_{k} = (\alpha_{k,\uparrow},
\beta_{k,\uparrow}, \alpha_{k,\downarrow},
\beta_{k,\downarrow})^{T}$, and $\mathbf{u}_{k}$ is the
cell-periodical eigenstate of the corresponding band, satisfying
$\hat{h}(k)\mathbf{u}_{k}  =E_{k}\mathbf{u}_{k}$. Follow
Ref.\cite{M. J. Rice}, we require that $\alpha_{k,\sigma} =
\alpha_{k+G,\sigma}$ and $\beta_{k,\sigma} = -\beta_{k+G,\sigma}$,
where $G=2\pi/d$ is the reciprocal lattice vector.

From Eq.(\ref{2}), we obtain the energy spectrums
\begin{eqnarray}
E_{k}= \pm \sqrt{|\rho_{k}|^{2} + |\delta_{k}|^{2} \pm
\sqrt{2(|\rho_{k}|^{2} |\delta_{k}|^{2} + \text{Re}(\rho_{k}^{2}
{\delta_{k}^{*}}^{2}))}}. \label{6}
\end{eqnarray}
We know the necessary condition for  topological phase transition
is the closure of the energy gap. When $\lambda = \lambda^{'} =
0$, the energy spectrums reduce to $E_{k}=\pm|\rho_{k}|$, it is
direct to find that the energy gap is closed only when $J=J^{'}$
at $k=\pi/d$. In Ref.\cite{M. Atala}, it is shown that when $J
> J^{'}$, each of the two occupied bands has $\varphi_{Zak} = \pi/2$,
and the system is topologically trivial, When $J<J^{'}$, each band
has $\varphi_{Zak} = -\pi/2$, and the system is topologically
nontrivial. To characterize the topology of Bloch bands
conveniently, we define a new topological invariant $\nu$,
\begin{eqnarray}
(-1)^{\nu} = sgn(\varphi_{Zak}). \label{7}
\end{eqnarray}
$\nu=0$ corresponds to a trivial band and $\nu=1$ corresponds to a
topologically nontrivial band. Based on this definition, the
phases of the system are classified by an integer defined as
$Z=\nu_{1}+\nu_{2}$, where $\nu_{1}$ is the topological invariant
of the lower occupied band, and $\nu_{2}$ is the topological
invariant of the upper occupied band. However, in the following we
use $Z(\nu_{1},\nu_{2})$, instead of $Z$, to characterize the
phases, because we can directly read the topology of every
occupied band from $Z(\nu_{1},\nu_{2})$.

\begin{figure}
\subfigure{\includegraphics[width=4cm, height=4cm]{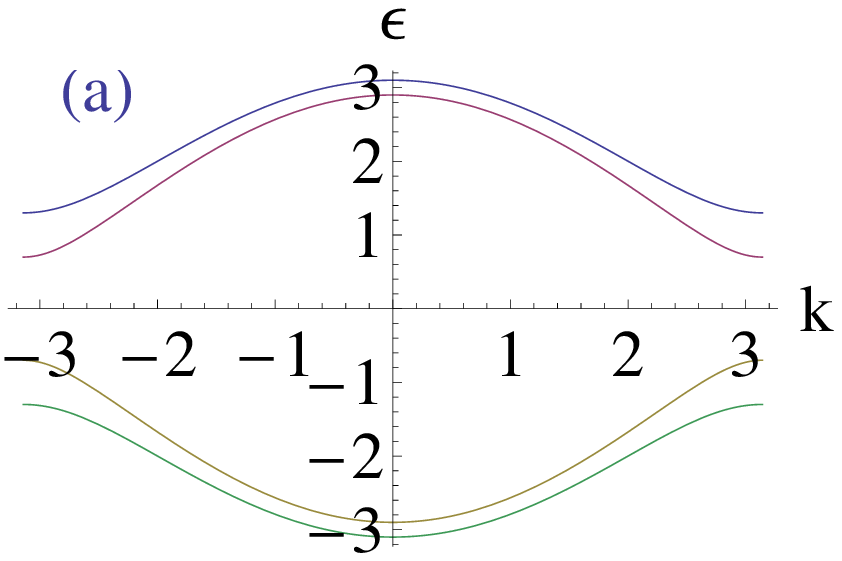}}
\subfigure{\includegraphics[width=4cm, height=4cm]{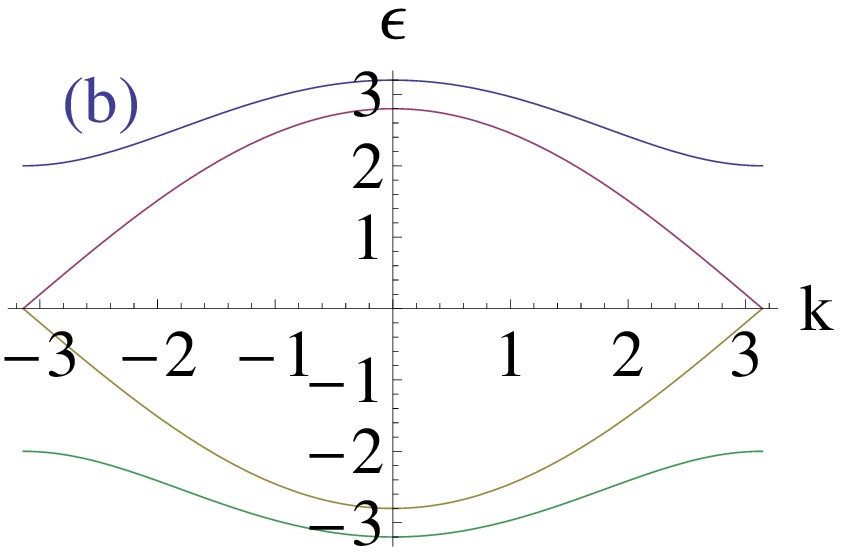}}
\subfigure{\includegraphics[width=4cm, height=4cm]{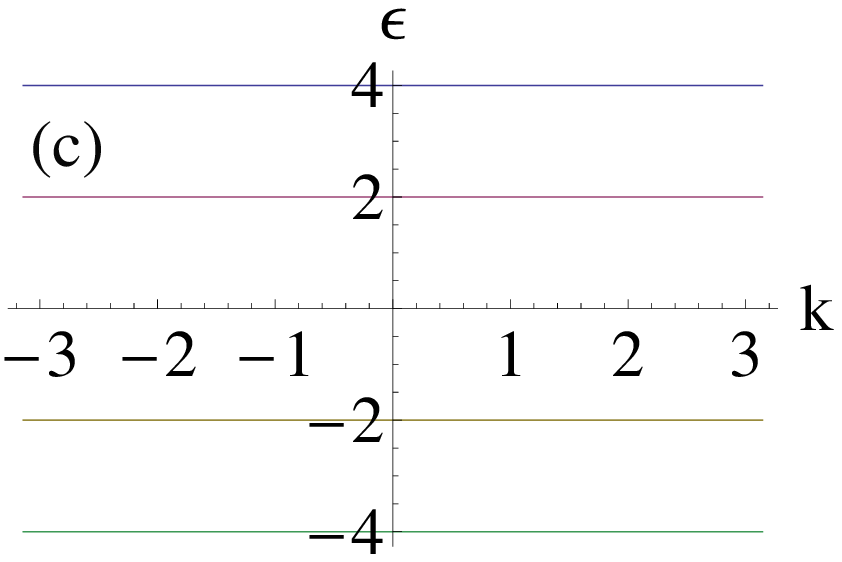}}
\subfigure{\includegraphics[width=4cm, height=4cm]{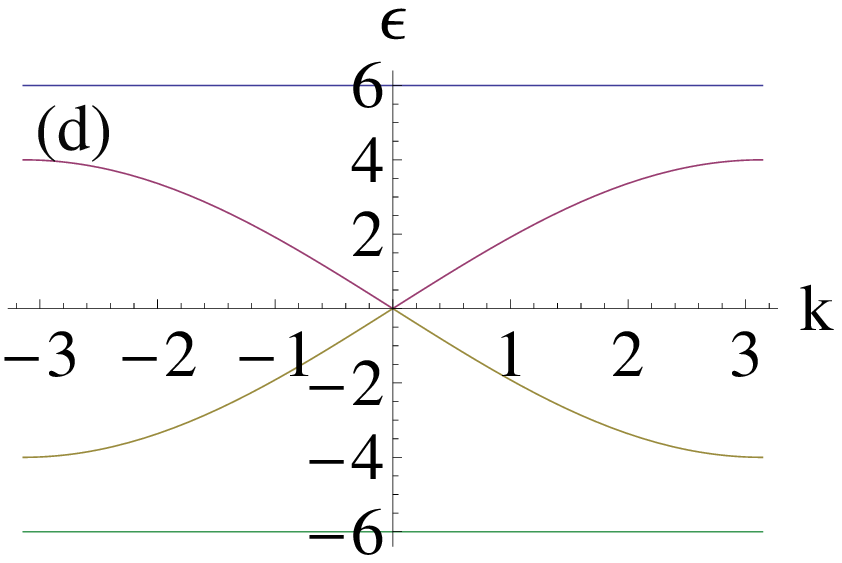}}
\caption{ (color online) Energy dispersion relation with parameters:
(a) $J=2$ $J^{'}=1$, $\lambda=0.2$, $\lambda^{'}=0.1$.
(b) $J=2$ $J^{'}=1$, $\lambda=0.6$, $\lambda^{'}=0.4$. (c) $J=2$ $J^{'}=1$, $\lambda=2$, $\lambda^{'}=1$.
(d) $J=2$ $J^{'}=1$, $\lambda=4$, $\lambda^{'}=1$.}\label{fig1}
\end{figure}

{\it Topological phase transitions induced by spin-orbit
coupling.---} When $\lambda=\lambda^{'}\neq0$, the degeneracy of
the bands is lifted, and only the energy gap between the upper
occupied band and the lower unoccupied band can be closed (see
Fig.\ref{fig1}(a)(b)(c)(d)). The condition for the closure of
energy gap is now modified. The new criterion is given by
\begin{eqnarray}
&&|\epsilon_{1}(k_{c})|=|\epsilon_{2}(k_{c})|,\nonumber\\
&&\theta_{1}(k_{c})-\theta_{2}(k_{c})=0 \quad (mod\,\pi).
\label{8}
\end{eqnarray}
A little investigation shows that the two conditions can only be
simultaneously satisfied when the parameters satisfy the relations
$(\lambda^{2}+{\lambda^{'}}^{2})-(J^{2} + {J^{'}}^{2}) = \pm 2
(\lambda \lambda^{'} + J J^{'})$.

For weak spin-orbit coupling ($\lambda^{2} + {\lambda^{'}}^{2} <<
J^{2} + {J^{'}}^{2}$), the Bloch bands do not change much and the
upper occupied band and the lower unoccupied band still touch at
$k_{c}=\pi/d$ (see Fig.\ref{1}(a)(b)). At the gap-closure point,
the parameters should satisfy the relation $(\lambda^{2} +
{\lambda^{'}}^{2}) - (J^{2} + {J^{'}}^{2}) = -2(\lambda
\lambda^{'} + J J^{'})$. If we take $\lambda = \lambda^{'} = 0$,
this relation just gives $J = J^{'}$, agreeing with the result
discussed above. From the criterion above, the parameter relation
for the closure of energy gap is given by
\begin{eqnarray}
|J - J^{'}| = \lambda + \lambda^{'}. \label{9}
\end{eqnarray}
For the sake of discussion, here we first assume $J > J^{'}$. This
assumption gives $J = J^{'} + \lambda + \lambda^{'}$. This simple
relation indicates that spin-orbit coupling changes the transition
point in a linear way.

For strong spin-orbit coupling ($\lambda^{2} + {\lambda^{'}}^{2}
> J^{2} + {J^{'}}^{2}$), the form of the Bloch bands has changed a lot
and the upper occupied band and the lower unoccupied band now
touch at $k_{c}=0$ (see Fig.\ref{1}(c)(d)). The criterion for gap
closure is $(\lambda^{2} + {\lambda^{'}}^{2}) - (J^{2} +
{J^{'}}^{2}) = 2(\lambda \lambda^{'} + J J^{'})$. The parameter
relation deduced from the criterion above is given by
\begin{eqnarray}
|\lambda - \lambda^{'}| = J + J^{'}. \label{10}
\end{eqnarray}

To confirm whether the gap closure discussed above corresponds to
a topological phase transition, we need to calculate the Zak phase
of the upper occupied band before and after the closure of gap. We
first consider the case with weak spin-orbit coupling. Before the
discussion, we remind the fact that only when energy gap gets
closed, the topology of the bands can change, and therefore, we
can choose special value for the parameters in every gapped region
to calculate the Zak phase. Before the closure of gap, based on
Eq.(\ref{9}), we choose $J=2$, $J^{'}=0$, $\lambda=1$,
$\lambda^{'}=0$, a few steps of calculation shows that the
four-component spinor $\mathbf{u}_{k}$ of the upper occupied Bloch
band takes the form,
\begin{eqnarray}
\mathbf{u}_{k}=\frac{1}{2}(1, e^{-ikd/2}, 1, e^{-ikd/2})^{T},\label{11}
\end{eqnarray}
then based on Eq.(\ref{5}), it is direct to obtain
$\varphi_{Zak}(J > J_{c}) = \pi/2$ (where $J_{c} = J^{'} + \lambda
+ \lambda^{'}$), or equivalently, $\nu_{2} = 0$. This indicates
the system is a trivial insulator, agreeing with the fact that the
parameters we choose above can be continued to the case with
parameters $\lambda = \lambda^{'} = 0$ and $J > J^{'}$, without
the closure of gap. After the closure of gap and the re-open of
gap, based on Eq.(\ref{9}), we choose Fig.\ref{fig1}(c)'s
parameters: $J=2$, $J^{'}=1$, $\lambda=2$, $\lambda^{'}=1$, then
$\mathbf{u}_{k}$ of the upper occupied Bloch band is given by
\begin{eqnarray}
\mathbf{u}_{k}=\frac{1}{2}(1, e^{ikd/2}, 1, e^{ikd/2})^{T},\label{12}
\end{eqnarray}
then we obtain $\varphi_{Zak}(J<J_{c})=-\pi/2$, or equivalently,
$\nu_{2}=1$. This indicates the upper occupied band which is
completely flat (see Fig.\ref{fig1}(c)) is of nontrivial topology.
If we keep $J=\lambda$, we find that varying $\lambda^{'}$ with
the constraint $\lambda^{'}<\lambda$ does not change the
completely flat form of the upper occupied band. This suggests
that the topological flat band can exist in a broad region.
Flat-bands are of great interest because they play an important
role in the study of strong correlated phenomena. One-dimensional
flat bands with nontrivial topology have already been considered
in models similar to the SSH model (only similar in form, the
underlying topology is different) and the authors have found the
existence of fractional topological phase \cite{H. Guo, Huaiming
Guo, J. C. Budich}.

$\nu_{2}=1$ indicates when $\lambda+\lambda^{'}>J-J^{'}$, the
system is driven into the topological phase $Z(0,1)$. As $J$ keeps
larger than $J^{'}$, the dimerization of the topological phase
$Z(0,1)$ is the same as the trivial phase $Z(0,0)$. This suggests
an important fact that  with the appearance of spin-orbit
coupling, using the simple picture of dimerization to directly
classify the phases is broken down.

Based on the analysis above, when $J=J_{c}$, accompanying the
closure of energy gap, a topological phase transition takes place,
with a change of Zak phase $\delta\varphi_{Zak}=\pi$ (mod $2\pi$).

With continuously increasing the strength of spin-orbit coupling,
the re-open gap will get closed again. In the region
$|\lambda-\lambda^{'}|>J+J^{'}$, we choose $J=1$, $J^{'}=0$,
$\lambda=2$, $\lambda^{'}=0$ to calculate the Zak phase. The
$\mathbf{u}_{k}$ of the upper occupied Bloch band is given by
\begin{eqnarray}
\mathbf{u}_{k}=\frac{1}{2}(-1, e^{-ikd/2}, -1, e^{-ikd/2})^{T},\label{13}
\end{eqnarray}
then the same procedure produces
$\varphi_{Zak}(|\lambda-\lambda^{'}|>J+J^{'})=\pi/2$, or
equivalently, $\nu_{2}=0$. The system return to the trivial phase
$Z(0,0)$. Therefore, the two closures of gap both correspond to
topological phase transition with a change of Zak phase
$\delta\varphi_{Zak}=\pi$ (mod $2\pi$).

Above, we have assumed $J>J^{'}$, if we instead assume $J^{'}>J$,
we find the gap closure also always correspond to topological
phase transition with a change of Zak phase
$\delta\varphi_{Zak}=\pi$ (mod $2\pi$). A big difference from the
case with $J > J^{'}$ is that, for $J^{'} > J$, the topological
phase transition is between two topological phases with different
topological invariants, and therefore, the system always hosts end
bound states. The only change is the number of the end bound
states.

\begin{figure}
\subfigure{\includegraphics[width=4cm, height=4cm]{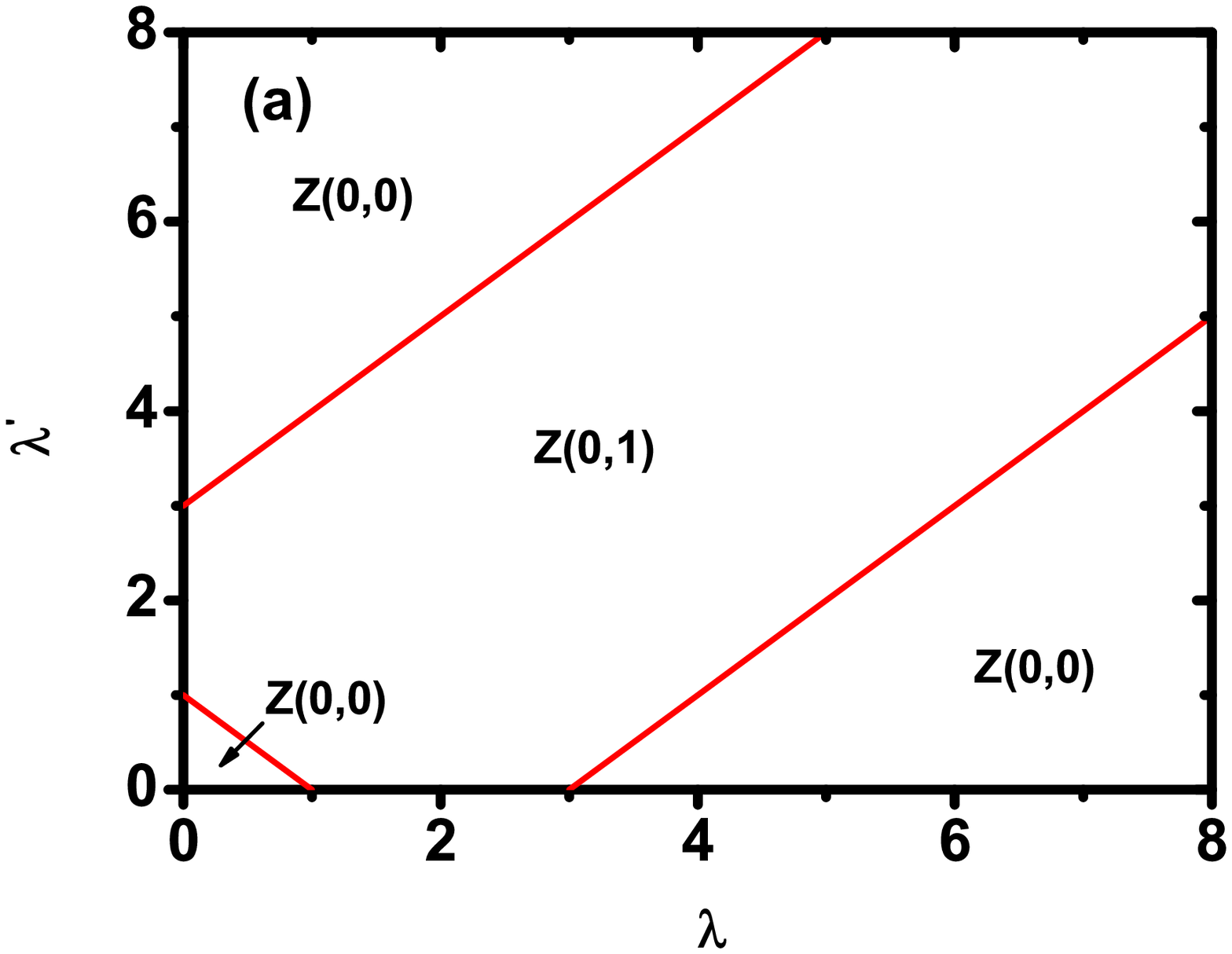}}
\subfigure{\includegraphics[width=4cm, height=4cm]{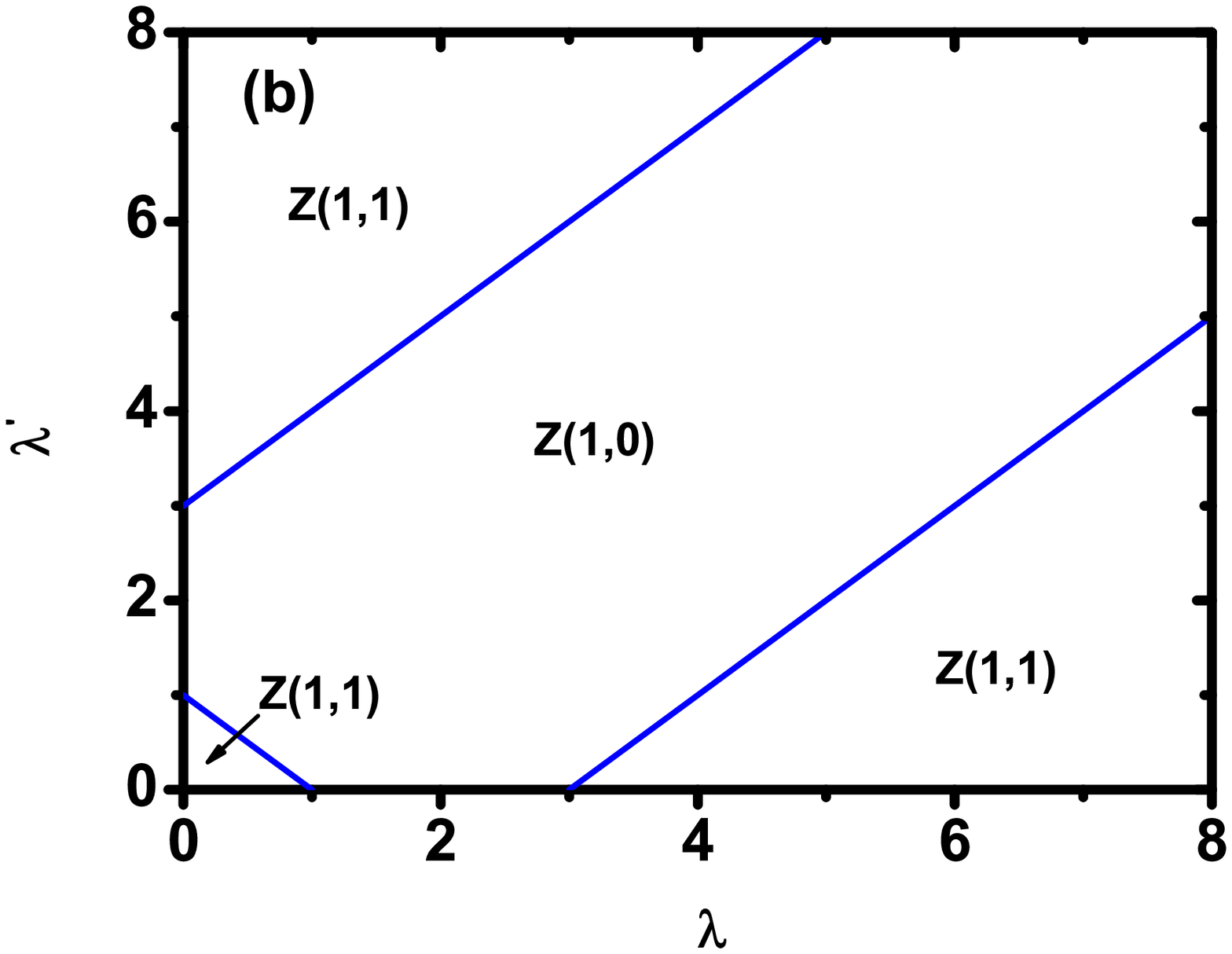}}
\subfigure{\includegraphics[width=4cm, height=4cm]{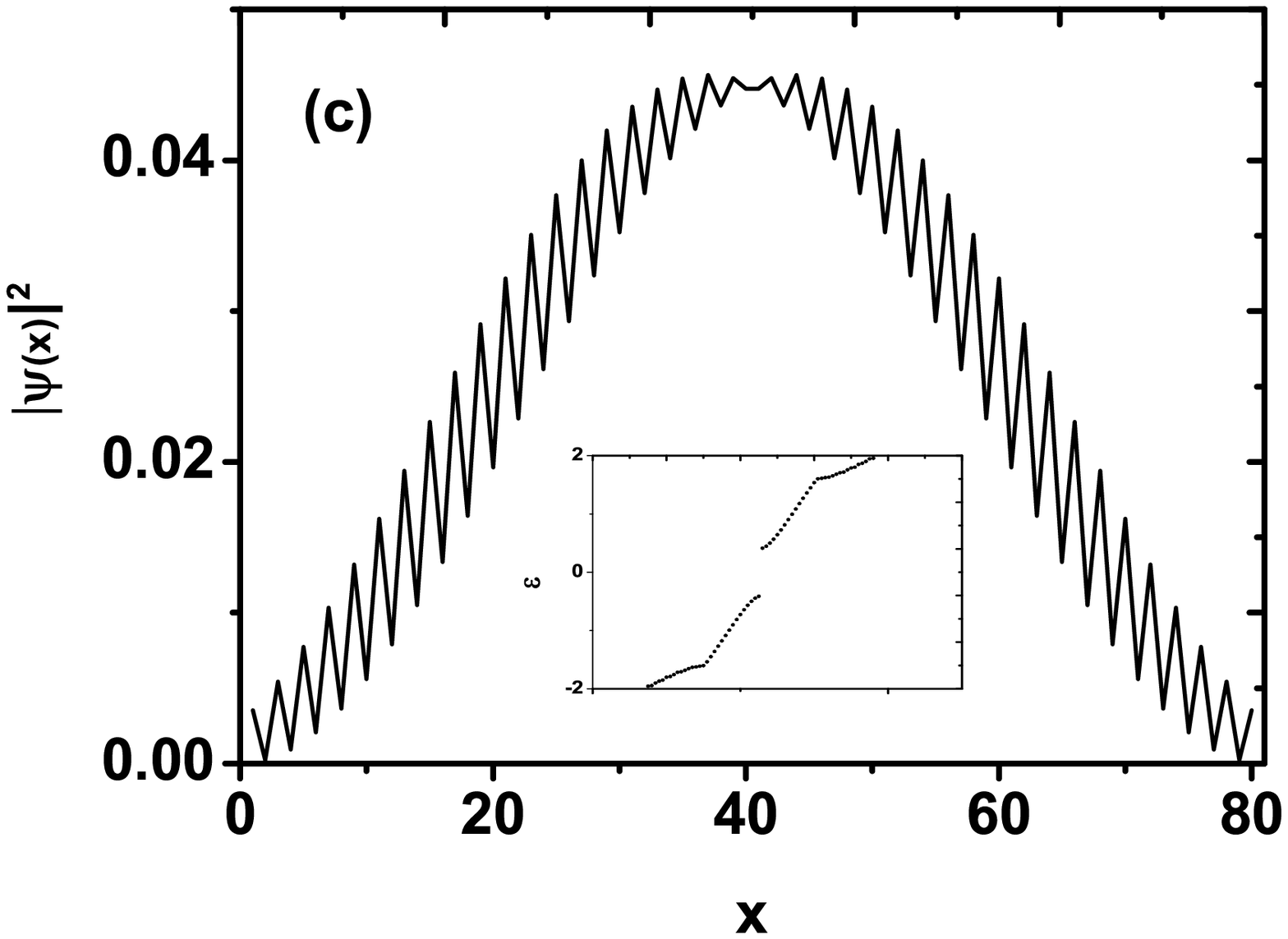}}
\subfigure{\includegraphics[width=4cm, height=4cm]{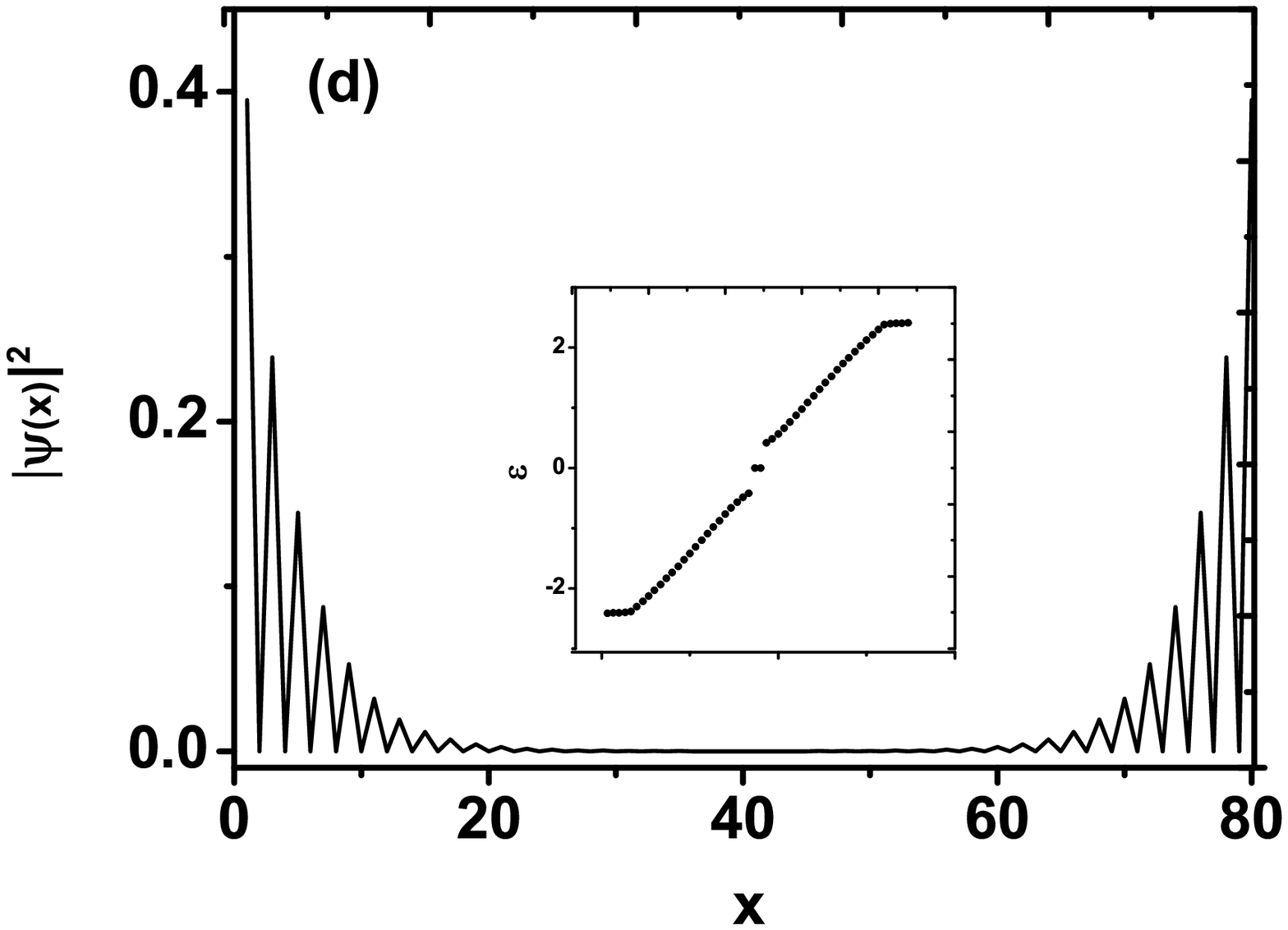}}
\subfigure{\includegraphics[width=4cm, height=4cm]{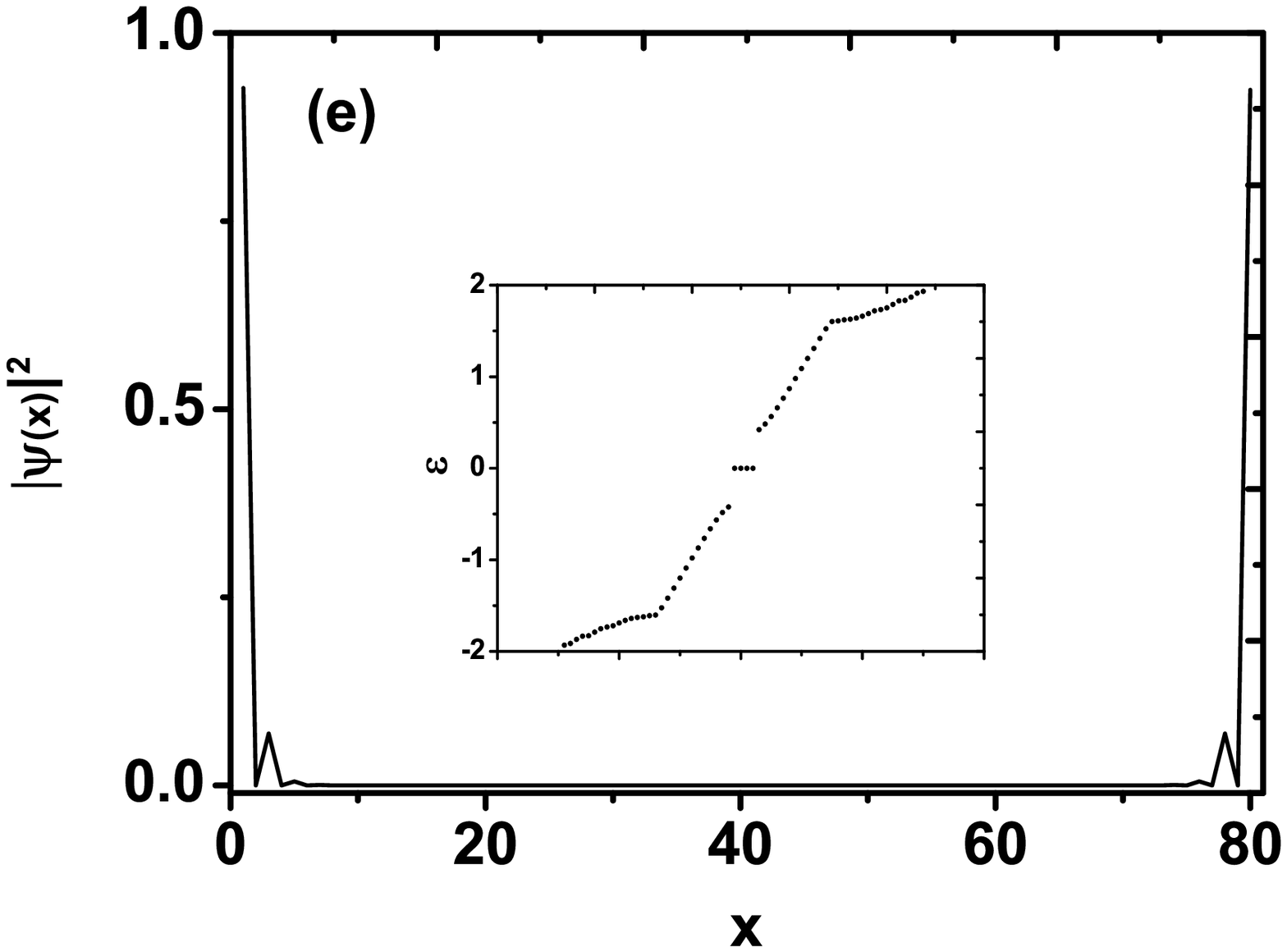}}
\subfigure{\includegraphics[width=4cm, height=4cm]{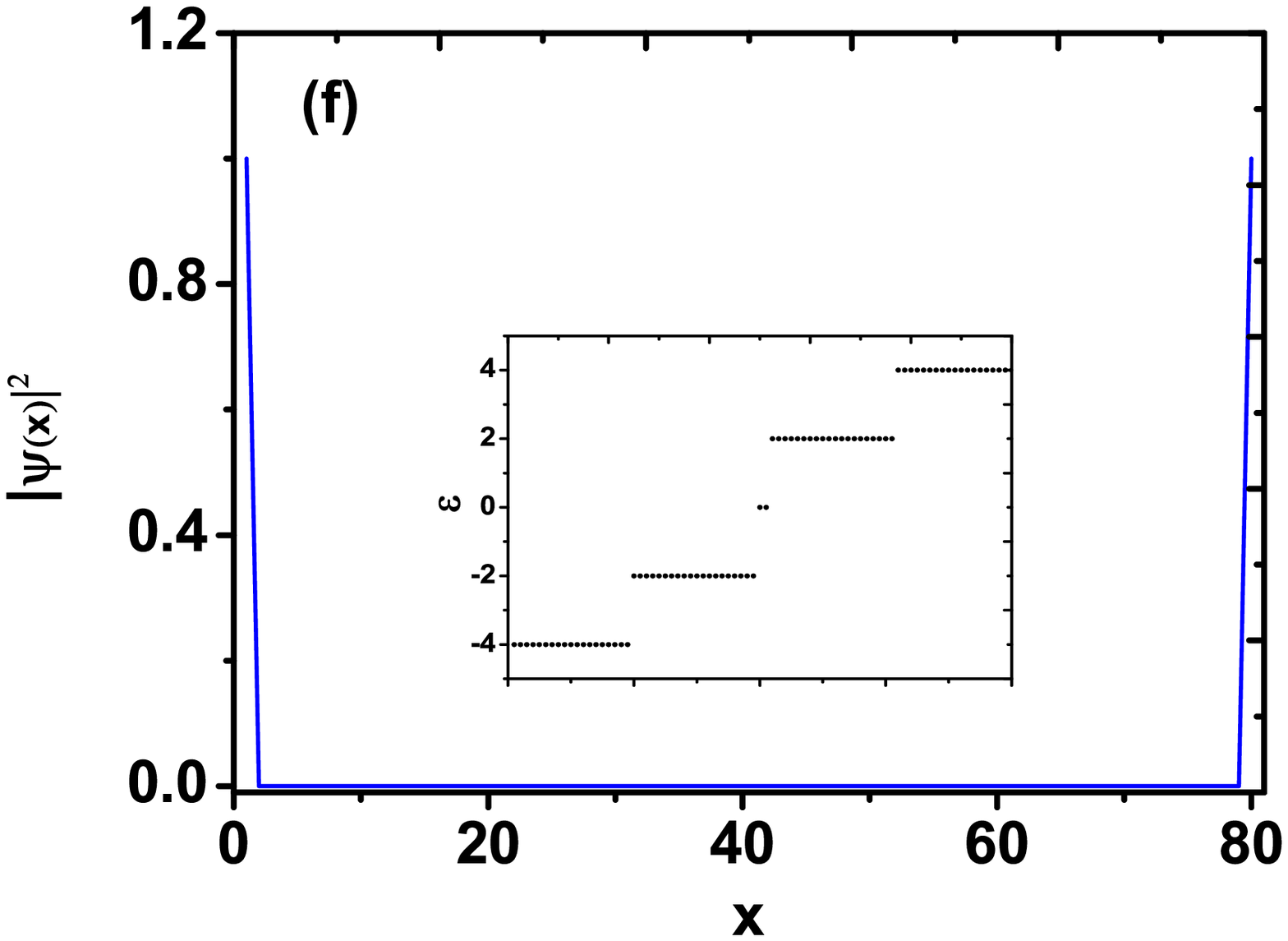}}
\caption{ (color online) (a)(b) are Phase diagrams, with parameters: (a) $J=2$ $J^{'}=1$.
(b) $J=1$ $J^{'}=2$. (c)(d)(e)(f) show the wave functions for excitations with the lowest energy with parameters, (c)
$J=2$, $J^{'}=1$, $\lambda=0.4$, $\lambda^{'}=0.2$,
(d) $J=2$, $J^{'}=1$, $\lambda=0.6$, $\lambda^{'}=0.8$. (e) $J=1$, $J^{'}=2$,  $\lambda=0.4$, $\lambda^{'}=0.2$.
(f) $J=1$, $J^{'}=2$, $\lambda=1$, $\lambda^{'}=2$. The insets in (c)(d)(e)(f) are the corresponding energy spectrums with $0$, $2$,
$4$, $2$ in-gap zero-energy bound states, respectively.}\label{fig2}
\end{figure}

Based on the analysis above, the phase diagrams can be directly
obtained. The phase diagrams for $J>J^{'}$ and $J<J^{'}$ have
similar form. From Fig.\ref{fig2}(a)((b)), we can see by
continuously increasing the strength and asymmetry of spin-orbit
coupling, the system first undergoes a topological phase
transition from the trivial phase $Z(0,0)$ (topological phase
$Z(1,1)$) to the topological phase $Z(0,1)$ $(Z(1,0))$ and then
undergoes another topological phase transition from the
topological phase $Z(0,1)$ $(Z(1,0))$ back to the trivial phase
$Z(0,0)$ (topological phase $Z(1,1)$). Fig.\ref{fig2}(c) shows
that in the trivial phase $Z(0,0)$, there is no zero energy bound
state at the end of the one-dimensional system.
Fig.\ref{fig2}(d)(f) show that the topological phase $Z(0,1)$ or
$Z(1,0)$ hosts single zero-energy bound state at each end.
Fig.\ref{fig2}(e) shows that the topological phase $Z(1,1)$ hosts
two zero-energy bound states at each end. The difference of the
number of edge states suggests that the topological phase $Z(0,1)$
or $Z(1,0)$ is a new topological phase. As the Hamiltonian belongs to
the symmetry class $BDI$, the single zero-energy end state
should be the same as the one found in Ref.\cite{J.
Klinovaja} and therefore exhibits non-Abelian statistics. Consequently,
the single zero-energy end state is similar to an unpaired Majorana fermion
\cite{Kitaev} and has the potential application
in topological quantum computation.

{\it The effect of spin-orbit coupling to topological
excitations.---} It is well-known that the most famous excitations
in the SSH model are solitons and antisolitons which are movable
\cite{W. P. Su}. They are the domain walls of the two topological
distinct phases with different dimerizations. The physics of such
domain walls is captured by the famous Jackiw-Rebbi model \cite{R.
Jackiw} and the TLM model \cite{H. Takayama}. Based on the
Eq.(\ref{1}), for weak spin-orbit coupling, the continuum
Hamiltonian is given by
\begin{eqnarray}
\hat{h}(x)=-iv\partial_{x}\sigma_{z}+\Delta(x)\sigma_{x}+[iv^{'}\partial_{x}\sigma_{z}-\Delta^{'}(x)\sigma_{x}]\tau_{x}, \label{14}
\end{eqnarray}
where $v=(J+J^{'})d/2$, $v^{'}=(\lambda-\lambda^{'})d/2$, and $\Delta(x)=J-J^{'}$, $\Delta^{'}(x)=\lambda+\lambda^{'}$ in the case of uniform dimerization.
By making a rotation of spin, $\tilde{h}(x)=e^{i\frac{\pi}{4}\tau_{y}}\hat{h}(x)e^{-i\frac{\pi}{4}\tau_{y}}$, we obtain
\begin{eqnarray}
\tilde{h}(x)=-iv\partial_{x}\sigma_{z}+\Delta(x)\sigma_{x}+[iv^{'}\partial_{x}\sigma_{z}-\Delta^{'}(x)\sigma_{x}]\tau_{z}. \label{15}
\end{eqnarray}
the Hamiltonian now can be decoupled to two parts corresponding to two different helicities,
\begin{eqnarray}
\tilde{h}_{+}(x)=-i(v+v^{'})\partial_{x}\sigma_{z}+(\Delta(x)+\Delta^{'}(x))\sigma_{x},\nonumber\\
\tilde{h}_{-}(x)=-i(v-v^{'})\partial_{x}\sigma_{z}+(\Delta(x)-\Delta^{'}(x))\sigma_{x}. \label{16}
\end{eqnarray}
For $J>J^{'}$, $\tilde{h}_{+}(x)$ always describes a trivial
phase, and  $\tilde{h}_{-}(x)$ may describe a trivial phase or a
topological phase depending on the sign of
$(\Delta(x)-\Delta^{'}(x))$. For  $J<J^{'}$, on the contrary,
$\tilde{h}_{-}(x)$ always describe a topological phase, and
$\tilde{h}_{+}(x)$ may describe a trivial phase or a topological
phase depending on the sign of $(\Delta(x)+\Delta^{'}(x))$. Based
on the fact that the system undergoes a topological phase
transition when $\Delta(x)-\Delta^{'}(x)=0$ for $J>J^{'}$ and
$\Delta(x)+\Delta^{'}(x)=0$ for  $J<J^{'}$, the parameter relation
for uniform relation in Eq.(\ref{9}) is reobtained. For strong
spin-orbit coupling, as the energy bands now touch at $k_{c}=0$,
therefore, the roles played by the kinetic term
$-iv\partial_{x}\sigma_{z}$ and the order parameter term
$\Delta(x)\sigma_{x}$ are exchanged. The discussion of the case
with strong spin-orbit coupling is straightforward, and we neglect
it here.

For non-uniform dimerization, $\tilde{h}_{+}(x)$
describes topological excitations with one of the helicities
labeled as ``$+$", and $\tilde{h}_{-}(x)$ describes topological
excitations with the other one labeled as ``$-$". Based on
Eq.(\ref{16}), the decay properties of the wave functions take the
form
\begin{eqnarray}
\varphi_{\pm}(x)\propto\exp(-\frac{|\Delta(x)\pm\Delta^{'}(x)||x|}{\tilde{v}}), \label{17}
\end{eqnarray}
here we have used the assumption $\Delta(x)=-\Delta(-x)=J-J^{'}>0$
for $x<0$, $\Delta^{'}(x)=\lambda+\lambda^{'}<J-J^{'}$ and
Dirichlet boundary condition for simplicity, and $\tilde{v}=v\pm
v^{'}$ for $\tilde{h}_{\pm}(x)$, respectively. From Eq.(\ref{17}),
it is found that when $\Delta^{'}(x)=0$, the topological
excitations have a symmetric structure, however, once
$\Delta^{'}(x)\neq0$, topological excitations with helicity ``$+$"
(``$-$") become more localized at the left (right) side and more
extended at the right (left) side of the domain.

Another important result induced by spin-orbit coupling is that by
increasing spin-orbit coupling to make
$\lambda+\lambda^{'}>|J-J^{'}|$ (but
$|\lambda-\lambda^{'}|<J+J^{'}$), although the dimerization
configuration keeps the same, the topological properties of both
sides are changed (as shown in Fig.\ref{fig2}(a)(b)). A direct
result of the simultaneous change is
the disappearance of the moving topological excitations. It is
direct to confirm this by noting that in this parameter region, the
sign of $(\Delta(x)+\Delta^{'}(x))$ is always positive and the
sign of $(\Delta(x)-\Delta^{'}(x))$ is always negative.
Unlike the SSH model where the existence of moving topological
excitations makes the system actually conducting and the end
states in fact unprotected, the disappearance of the moving
topological excitations indicates the system turns to be a
real insulator in bulk, and consequently, the system is a
real topological insulator with well-protected end states. The
disappearance of the moving topological excitations also suggest
the two topological phases denoted by $Z(0,1)$ and $Z(1,0)$ are
the same, agreeing with the fact that the Hamiltonian is classified by
an integer. Base on Fig.\ref{fig2}(a)(b), topological excitations
will reappear by further increasing the strength and the asymmetry
of spin-orbit coupling.

{\it Conclusion.---} The introduction of spin-orbit coupling
greatly riches the physics of SSH model. First, with the lift of
the degeneracy, we find the usual $Z_{2}$ invariant classifying
the phases of SSH model can no longer fully classify all phases.
The new topological phase corresponding to $Z=1$ hosts interesting
single zero-energy bound state which exhibits non-Abelian statistics
at each end and is stable against the variation of
dimerization, therefore, it can have great potential application in
topological quantum computation. Second, in the region of
topological phase $Z=1$, completely flat band with nontrivial
topology can be formed by tuning the spin-orbit coupling. Third,
for the case with non-uniform dimerization, spin-orbit coupling in
weak region changes the symmetrical form of topological
excitations, and with increasing the strength and the asymmetry of
spin-orbit coupling, the topological excitations will disappear
when a topological phase transition takes place and then re-appear
when another topological phase transition takes place. These
results suggest spin-orbit coupling can be used to control the
phases and the topological excitations of the system. The progress
on experiments makes the observation of the new topological phase
and the topological phase transitions induced by spin-orbit
coupling in experiments within current ability.

{\it Acknowledgments.---} We would like to thank Xiaosen Yang for
helpful discussions. This work is supported by NSFC. Grant
No.11275180.

\end{document}